\documentstyle[12pt]{article}
\newcommand{\be}{\begin{equation}}
\newcommand{\ee}{\end{equation}}
\newcommand{\bea}{\begin{eqnarray}}
\newcommand{\eea}{\end{eqnarray}}
\newcommand{\no}{\nonumber\\}
\def\r{{\bf r}}
\def\R{{\bf R}}
\def\p{{\bf p}}
\def\k{{\bf k}}
\def\x{{\bf x}}
\def\y{{\bf y}}
\def\z{{\bf z}}

\input BoxedEPS
\SetRokickiEPSFSpecial %%for unix dvips
\HideDisplacementBoxes 

\begin{document}
\hfill CTP \# 2561

\hfill August, 1996
\vskip 0.25in
\centerline{\large Bose Condensate in External Potential:}
\centerline{\large A Thomas-Fermi Approach\footnote
{Contribution to Workshop on Bose-Einstein Condensation, Institute for
Theoretical Atomic and Molecular Physics, Harvard-Smithsonian Center
for Astrophysics, Cambridge, MA, (August 19-30, 1996).}}
\vskip 0.1in
\centerline{Kerson Huang}
\vskip 0.1in
\centerline{Center for Theoretical Physics, Laboratory for
Nuclear Science}
\centerline{       and Department of Physics}
\centerline{   Massachusetts Institute of Technology}
\centerline{   Cambridge, MA 02139}

\vskip 0.5in

\section{Thomas-Fermi Theory}

The Thomas-Fermi method \cite{T} \cite{F}
was designed for the calculation of
the electron density in a heavy atom, by treating the electrons 
as locally free. Lieb and Simon \cite{LS}
showed that the treatment is exact in the limit when 
the atomic number goes to infinity. 
Application to a confined Bose condensate was pioneered
by Goldman, Silvera, and Legget \cite{GSL}, and by Oliva \cite{O},  
and recently reconsidered by Chou, Yang, and Yu \cite{CYY}.
I shall describe some work on this subject, done in collaboration with 
E. Timmermans and P. Tommasini \cite{TTH}.

First, let us review the original method of Thomas and Fermi. Suppose
$V(r)=-e\Phi(r)$ denotes the effective potential energy of an electron 
in an atom, at a distance $r$  from the nucleus. (See Fig.1). 

\medskip
\begin{figure}[htbp]
\centerline{\BoxedEPSF{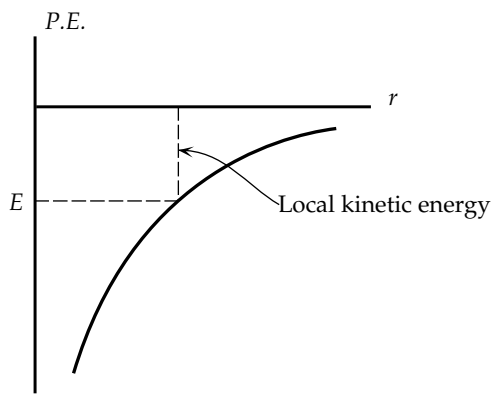 scaled 500}}
\caption{Potential energy of an electron in atom.}
\end{figure}
\medskip

\noindent The condition that
the electron is in a bound orbit is that
\be
{\hbar^2 k^2\over 2m} +V(r) \le 0
\ee
where $k$ is the local wave vector. Assume that all available states
are occupied. Then the local Fermi momentum $\hbar k_F(r)$ is  given by
\be
{\hbar^2 k^2_F(r)\over 2m} +V(r) = 0
\ee
which gives
\be
\hbar k_F=\sqrt{2me\Phi(r)}
\ee
We know that $k_F(r)$ is related to tke local  density $n(r)$ by
\be
2(4\pi/3)k_F^3=n(r)
\ee
where the factor 2 comes  from spin. This relates the density to
the potential $\Phi(r)$. We now use the Poisson equation
\be
\nabla^2 \Phi(r) =-4\pi[Ze\delta({\bf r}) -e n(r)]
\ee
For $r\ne 0$ this equation is of the form
\be
\nabla^2\Phi(r) + C \Phi^{3/2}(r)=0
\ee
where $C$ is a constant. One can solve this equation, and then obtain
$n(r)$. A comparison between Thomas-Fermi and Hartree results 
for Rb is sketched in Fig.2. 

\medskip
\begin{figure}[htbp]
\centerline{\BoxedEPSF{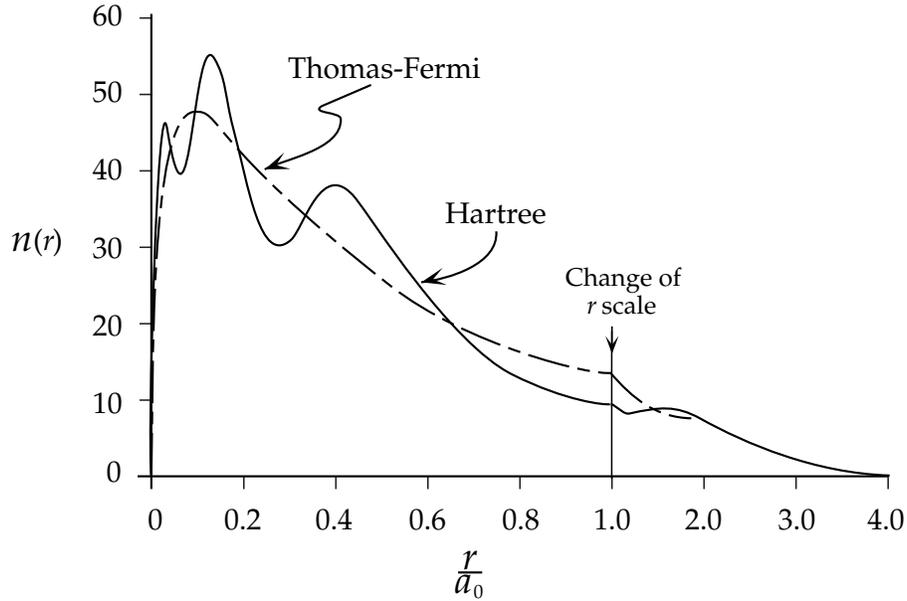 scaled 800}}
\caption{Electron density in Rb: Comparison between Thomas-Fermi and 
Hartree approximations.}
\end{figure}
\medskip

The essence of the method is that one assumes there is a local chemical 
potential $\mu_{\rm eff}(r)$, related to the true chemical potential $\mu$
by
\be
\mu_{\rm eff}(r) = \mu-V(r)
\ee
In the earlier discussion, the chemical potential (Fermi energy) was
taken to be zero. 

\section{Ideal Bose Gas in External Potential}

Can we apply this idea to a Bose gas? Let  us first concsider
 an ideal Bose gas in an external potential, at a temperature $T$
 above the transition point. Take the potential to be harmonic:
\be
V(r) = {1\over 2}m\omega^2 r^2
\ee
For the Thomas-Fermi idea to be applicable, we require
\be
{\hbar\omega\over kT}\ll1
\ee
Above the transition temperature, the density is related to 
the fugacity $z=\exp(-\mu/kT)$ through
\be
n={1\over \lambda^3} g_{3/2}(z)
\ee
where $\lambda=\sqrt{2\pi\hbar^2/mkT}$ is the thermal wavelength, and
\be
g_{3/2}(z)=\sum_{\ell=1}^\infty {z^\ell\over \ell^{3/2}}
\ee
This suggests that in the presence of an external potential we 
take the local density to be
\be
n(r) = {1\over \lambda^3} g_{3/2}\left(z e^{-\beta V(r)}\right)
\ee
where $\beta=1/kT$. Integrating both sides over all space, we obtain
an expression for the total number of particles:
\be
N = {1\over \lambda^3} \int d^3r g_{3/2}\left(z e^{-\beta V(r)}\right)
\ee
We know that $g_{3/2}(z)$ is bounded for $0\le z\le 1$. So the right
side is bounded. This forces Bose-Einstein  condensation when $N$
exceeds the bound. The number of atoms in the condensate $N_0$ is
given through
\be
N = N_0 + {1\over \lambda^3} \int d^3r g_{3/2}\left(z e^{-\beta V(r)}\right)
\ee
In this intuitive approach, however, the Bose condensate was not described 
accurately. 

As it turns out, the problem can be solved exactly [5]:
\be
n(r) = {z_1\over 1-z_1}|\psi_0(r)|^2+{1\over\lambda^3}G(z,r)
\ee
where $\psi_0(r)$ is the  ground-state wave function in the potential, 
and
\be
G(z_1,r) = {1\over (2\epsilon)^{3/2}}\sum_{\ell=1}^\infty z_1^\ell
\left\{ {\exp[-(r/r_0)^2 \tanh(\epsilon\ell/2)]\over 
[1-\exp(-2\epsilon\ell)]^{3/2} }
-\exp[-(r/r_0)^2] \right\}
\ee
with
\bea
z_1 &=& z e^{-3\hbar\omega/2kT}\no
\epsilon &=& \hbar\omega/kT\no
r_0&=&\sqrt{kT/2\pi m\omega^2}
\eea
The explicit occurence of $\psi_0(r)$ shows that Bose condensation 
occurs in the ground state $\psi_0$ of the potential. The zero-momentum state
is irrelevant here. The total number of particles is
\be
N = {z_1\over 1-z_1}+{1\over\lambda^3}\int d^3r G(z,r)
\ee
which shows that the condensation is a continuous process, though 
it may appear to be abrupt, when $N$ is so large that the first
term can be neglected except near $z_1=1$.

The Thomas-Fermi approximation is good when $\epsilon \ll 1$. In that case
we have 
\be
G(z_1,r)\approx g_{3/2}\left( z_1 e^{-\beta V(r)}\right)
\ee
and therefore
\be
n(r) \approx {z_1\over 1-z_1}|\psi_0(r)|^2+{1\over\lambda^3}
g_{3/2}\left( z_1 e^{-\beta V(r)}\right)
\ee
which is similar to the naive formula, except for a better representation
of the condensate. (The replacement of $z$ by $z_1$ is inconsequential.)
The lesson is that a purely intuitive approach is not satisfactory, and we
need a systematic method.

\section{Uniform Dilute Interacting Bose Gas}

The underlying idea of the Thomas-Fermi approach is to treat
a nonuniform condensate as locally uniform, with a slowly varying
density. I will first review the properties of a uniform Bose gas
in the dilute limit, with interparticle interactions taken into
account through a scattering length $a\ge 0$.

The annihilation operator $a_k$ of a particle of 
momentum $\hbar{\bf k}$ satisfies the commutation relation
\be
[a_k, a_{k'}^\dagger]=\delta_{k k'}
\ee
We make a Bogolubov transformation to quasiparticle operators $\eta_k$:
\be
a_k=x_k \eta_k -y_k \eta_{-k}^\dagger \quad ({\bf k}\ne 0)
\ee
and require that the transformation be canonical, {\it i.e.}
\be
[\eta_k, \eta_{k'}^\dagger]=\delta_{k k'}
\ee
This leads to the condition
\be
x_k^2-y_k^2=1
\ee
which can be satisfied by putting
\bea
x_k &=& \cosh\sigma_k\no
y_k &=& \sinh \sigma_k
\eea
This a convenient parametrization, because interesting quantities find
simple expressions:
\bea
\rho_k&\equiv&\langle a_k^\dagger a_k\rangle 
={1\over 2}[\cosh (2\sigma_k) -1]\no
\Delta_k&\equiv& -\langle a_k a_{-k}\rangle ={1\over 2}\sinh(2\sigma_k)
\eea
where $\rho_k$ measures the depletion of the unperturbed condensate:
\be
N_0=N-\sum_{{\bf k}\ne 0}\rho_k
\ee
and $\Delta_k$, is a measure of off-diagonal long range order.

In the Bogolubov method, the annihilation operator for the 
zero-momentum state $a_0$ is equated to the c-number $\sqrt{N}$. 
Explicit solution of the problem gives
\be
\tanh(2\sigma_k)={\mu\over (\hbar^2 k^2/2m) +\mu}
\ee
where $\mu$ is the chemical potential:
\be
\mu={4\pi a\hbar^2n\over m}\left(1+{32\over 3}\sqrt{na^3\over \pi}\right)
\ee
with $n$ the particle density. Note that this cannot be 
continued to negative $a$; apparently, new physics arises when the 
scattering length turns negative. The excitation energy of a quasiparticle 
of momentum ${\bf p}$ is given by
\be
\epsilon_p =\sqrt{\left({p^2\over 2m} +\mu\right)^2-\mu^2}
\ee

\section{Quasiparticle Field Operator}

In the uniform case, the field operator $\Psi(\r)$ can be put in the form
\be
\Psi(\r)=a_0+\psi(\r)
\ee
with
\be
\psi(\r)=\Omega^{-1/2}\sum_{\k\ne 0} a_k e^{i\k\cdot\r}
\ee
where $\Omega$ is the spatial volume. We have
\be
[\psi(\r),\psi^\dagger(\r')]=\delta(\r-\r')
\ee
since $a_0$ is treated as a c-number. We can introduce a quasiparticle 
field operator:
\be
\xi(\r)\equiv\Omega^{-1/2}\sum_{\k\ne 0} \eta_k e^{i\k\cdot\r}
\ee
Note that the relation between $\psi$ and $\xi$ is non-local:
\be
\psi(\r)= \int d^3 y [X(\x-\y)\xi(\y)-Y^\ast(\x-\y)\xi^\dagger(\y)]
\ee 
where
\bea
X(\x-\y) &=& \Omega^{-1/2}\sum_{\k\ne 0} x_k e^{i\k\cdot(\x-\y)}\no
Y(\x-\y) &=& \Omega^{-1/2}\sum_{\k\ne 0} y_k e^{i\k\cdot(\x-\y)}
\eea

For a non-uniform Bose gas, we write 
\be
\Psi(\r)=\phi(\r) +\psi(\r)
\ee
where $\phi(\r)$ is a c-number function, such that
\be
\langle\psi(\r)\rangle =0
\ee
where $\langle\rangle$ denotes ground state expectation value. We
transform to quasiparticle operators by putting
\be
\psi(\r)= \int d^3 y [X(\x,\y)\xi(\y)-Y^\ast(\x,\y)\xi^\dagger(\y)]
\ee 
The requirement
\be
[\xi(\r),\xi^\dagger(\r')]=\delta(\r-\r')
\ee
leads to the condition
\be
\int d^3z [X(\x,\z) X(\z,\y)-Y(\x,\z) Y(\z,\y)]=\delta(\x-\y)
\ee
The fulfillment of this condition in a simple fashion will guide our
formulation of the Thomas-Fermi approximation.

\section{Wigner Representation}

In quantum mechanics, the Wigner distribution associated with a 
wave funcition $\psi(\r)$ is defined by
\be
\rho_W(\R,\p)\equiv\int d^3r \psi^\ast(\R+\r/2)\psi(\R-\r/2)e^{i\p\cdot\r/\hbar}
\ee
That is, we take the off-diagonal density at two different
points in space, and Fourier analyze with respect to the
relative separation. This is illustrated in Fig.3. 

\medskip
\begin{figure}[htbp]
\centerline{\BoxedEPSF{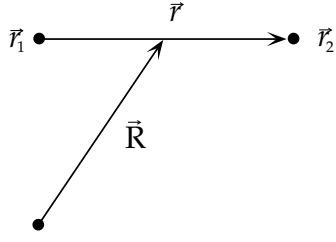  scaled 650}}
\caption{To get the Wigner distribution, Fourier analyze with respect
to relative coordinate.}
\end{figure}
\medskip

\noindent The Wigner distribution
is not positive-definite, and hence not a probability; but it
acts as a quasi-distribution function in phase space:
\bea
(\psi,f\psi)&\equiv&\int d^3r \psi^\ast(\r) f(\r) \psi(\r)=
\int{d^3R d^3p\over h^3} f(\R)\rho_W(\R,\p)\no
(\psi,\p\psi)&\equiv&\int d^3 r \psi^\ast(\r) {\hbar\over i}\nabla\psi(\r)=
\int{d^3R d^3p\over h^3} {\hbar\over i}\nabla\rho_W(\R,\p)
\eea

For a function $X(\x,\y)$ that depends on two spatial points,
we define its Wigner transform as
\be
X_W(\R,\p)\equiv\int d^3r X(\R+\r/2, \R-\r/2) e^{i\p\cdot\r}
\ee
with the inverse transform
\be
X(\x,\y)=\int{d^3p\over (2\pi)^3}e^{-i\p\cdot (\x-\y)}X_W((\x+\y)/2,\p)
\ee
If $C(\x,\y)$ has the form
\be
C(\x,\y)=\int d^3z A(\x,\z) B(\z,\y)
\ee
then its Wigner transform takes the form
\bea
C_W(\R,\p)&=&A_W(\R,\p) B_W(\R,\p) + 
{1\over 2i}\sum_{j=1}^3
\left( {\partial A_W\over\partial R_j}{\partial B_W\over\partial p_j}
-{\partial B_W\over\partial R_j}{\partial A_W\over\partial p_j}\right)\no
&&-{1\over 8}\sum_{j=1}^3
\left( {\partial^2 A_W\over\partial R_j^2}{\partial^2 B_W\over\partial p_j^2}
+{\partial^2 B_W\over\partial R_j^2}{\partial^2 A_W\over\partial p_j^2}
-2{\partial^2 A_W\over\partial R_j\partial p_j}
{\partial^2 B_W\over\partial R_j\partial p_j^2}\right)\no
&&+\cdots
\eea
The second term is the classical
Poisson bracket $\{A_W,B_W\}_{\rm PB}$. It and the subsequent terms all
depend on spatial derivatives, and would be small if the system is 
nearly unform. Thus our version of
 Thomas-Fermi approximation consists of
keeping only the first term. Errors incurred
can be estimated by calculating the next non-vanishing term.

In terms of the Wigner transform, we can write
\be
\int d^3z X(\x,\z) X(\z,\y)=X_W(\R,\p)X_W(\R,\p)
+{1\over 2i}\{X_W,X_W\}_{\rm PB}+\cdots
\ee
where the second terms vanishes identically. Thus,
for such an integral, errors incurred in  using the 
Thomas-Fermi approximation starts with subsequent terms.
The condition on $X$ and $Y$ therefore reads
\be
X_W^2(\R,\p)-Y_W^2(\R,\p)\approx 1
\ee
and is solved by setting
\bea
X_W(\R,\p)&=&\cosh\sigma(\R,\p)\no
Y_W(\R,\p)&=&\sinh\sigma(\R,\p)
\eea
This make the problem very similar to the uniform case. 

At zero temperature, the criterion for the validity of
the Thomas-Fermi approximation is
\be
\hbar\omega/ \mu\ll1
\ee
where $\hbar\omega$ is the characteristic energy of the external
potenial, and $\mu$ is the chemical potential. For the dilute interacting
Bose gas, $\mu$ is of order of the scattering length. Thus, the Thomas-Fermi
approximation can be used only when there are interparticle interactions.

\section{Variational Calculation}

We study the system defined by the Hamiltonian $H$, with
\be
H-\mu N = \int d^3x \Psi^\dagger h\Psi +{1\over 2}
\int d^3x d^3y \Psi^\dagger\Psi^\dagger V\Psi\Psi
\ee
where $V(\x)$ is the interparticle potential, and
\be
h=-{\hbar^2\over 2m}\nabla^2 + V_{\rm ext}(\x) -\mu
\ee
with $V_{\rm ext}(\x)$ the external potential. The ground state free 
energy is
\be
F=\langle H-\mu N\rangle
\ee
where $\langle\rangle$ means expectation value with respect to 
the ground state of $H-\mu N$. As mentioned before, we displace the field
by writing $\Psi=\phi+\psi$, where $\phi$ is a c-number function,
such that $\langle\psi\rangle=0$.
We assume a trial form for the ground state, so that $\langle F\rangle$ 
has the same form as in mean-field theory, {\it i.e.}, we can put
\be
\langle\psi^\dagger(\y)\psi^\dagger(\x)\psi(\x)\psi(\y)\rangle 
=\Delta^\ast(\y,\x)\Delta(\x,\y)+\rho(\y,\x)\rho(\x,\y)+\rho(\y,\y)\rho(\x,\x)
\ee
where
\bea
\rho(\x,\y)&=&\langle\psi^\dagger(\x)\psi(\y)\rangle\no
\Delta(\x,\y)&=&-\langle\psi(\x)\psi(\y)\rangle
\eea
The ground state free energy $F[\phi,\rho,\Delta;\mu]$ is  
a functional of $\phi$, $\rho$, and $\Delta$, and also depends on $\mu$ 
as a parameter. The requirement $\langle\psi\rangle=0$ means that there
are no terms in $F$ linear in $\phi$.

Although we do not need the trial state explicitly,
it can be explicitly constructed if desired. One can show that the wave 
functional of this state is of Gaussian form \cite{HT}. Thus, we have a true 
variational problem. 

We rewrite the functions $\rho$ and $\Delta$ in $F[\phi,\rho,\Delta;\mu]$
in terms of their Wigner transforms, and  implement our
version of the Thomas-Fermi approximation, as explained before.
We transform to quasiparticle field operators, and find that, 
as in the uniform case, $\rho_W$ and $\Delta$ are parametrized by a
single function:
\bea
\rho_W(\R,\p)&=&{1\over 2}[\cosh(2\sigma(\R,\p))-1]\no
\Delta_W(\R,\p)&=&{1\over 2}\sinh(2\sigma(\R,\p))
\eea
The Free energy reduces to the form $F=\int d^3R f(\R)$. 
We obtain equations for $\sigma$ and $\phi$ 
by minimizing $F$. The equation for $\phi$ is a modified 
Gross-Pitaevskii or non-linear Schr\"odinger equation (NLSE):
\be
\left[-{\hbar^2\over 2m}\nabla^2 +V_{\rm ext}(\r)+U(\r)-\mu
+v(0)\phi^2(\r)\right]\phi(\r)=0
\ee
where $U(\r)$ is a self-consistent potential that depends on $\sigma$.
It is unimportant for low densities.

\section{Dilute Interacting Gas in Harmonic Trap}

I will just quote some results for a dilute gas in a harmonic trap. The
external potential is
\be
V_{\rm ext}={\hbar\omega\over 2}\left(r\over L\right)^2
\ee
For particles of mass $m$,
\be
L =\sqrt{\hbar/m\omega}
\ee
which is the extend of the ground state wave function.
For the interparticle interation, we use a pseudopotential 
\be
{4\pi a\hbar^2\over m}\delta(\r){\partial\over \partial r}r
\ee
The sole effect of the differential operator above is
the removal of a divergence in the ground 
state energy. The three important lengths in the problem are
\bea
L&\quad&\mbox{(Extend of ground state wave function)}\no
a&\quad&\mbox{(Scattering length)}\no
R_0&\quad&\mbox{(Extend of condensate)}
\eea
They are illustrated in Fig.4.

\medskip
\begin{figure}[htbp]
\centerline{\BoxedEPSF{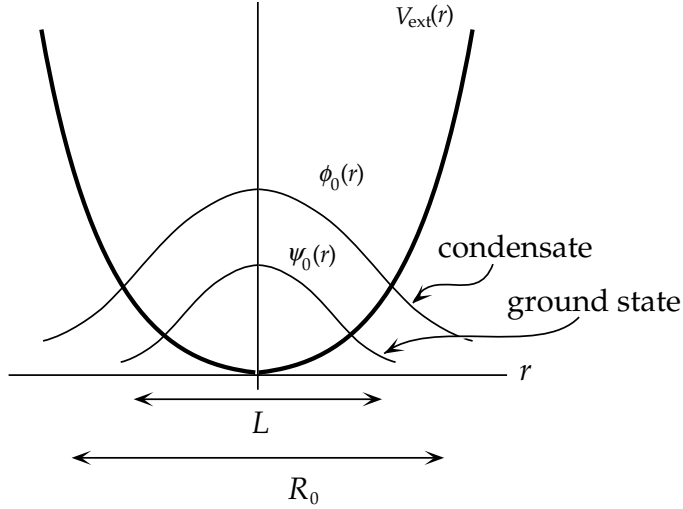 scaled 650}}
\caption{Length scales in atomic trap. Groundstate wave function is
$\psi_0(r)$. Condensate wave function is $\phi_0(r)$.}
\end{figure}
\medskip

For low densities, the non-linear coupled equations for $\sigma$ and 
$\phi$ are solved by iteration, and one iteration suffices.
The chemical potential is found to be
\be
\mu={\hbar\omega\over 2}\left(15 a\over 2L\right)^{2/5}
\left[1+{\sqrt{2}\over 60}\left(15a\over L\right)^{6/5} N^{1/5}\right]
\ee
The requirement $\hbar\omega/\mu\ll1$ means
\be
{L\over a N}\ll1
\ee
The extend of the condensate is given by
\be
{R_0\over L}=\left(15aN\over L\right)^{1/5}
\ee
For the method to be valid we must have $R_0>>L$.
By neglecting the term $\nabla^2\phi$ in the NLSE, we find
\be
\phi^2(r)={R_0^2\over 8\pi aL^4}\left[1-\left(r\over R_0\right)^2 \right]
\ee

In Fig.5 we show the shape of the condensate and estimated errors.
Fig.5(a) shows $\phi(r)$ as a function of $r$ in units of $L$, for
N=$10^3$, and $10^6$. 

Fig.5(b) shows the errors arising from the
neglect of $\nabla^2\phi$. This is ``trivial,'' as
it can be corrected through numerical computation.

Fig.5(c) shows the errors incurred due to the Thomas-Fermi approximation,
and are intrinsic to the method. They are small except at the edge of 
the condensate.

\medskip
\begin{figure}[htbp]
\centerline{\BoxedEPSF{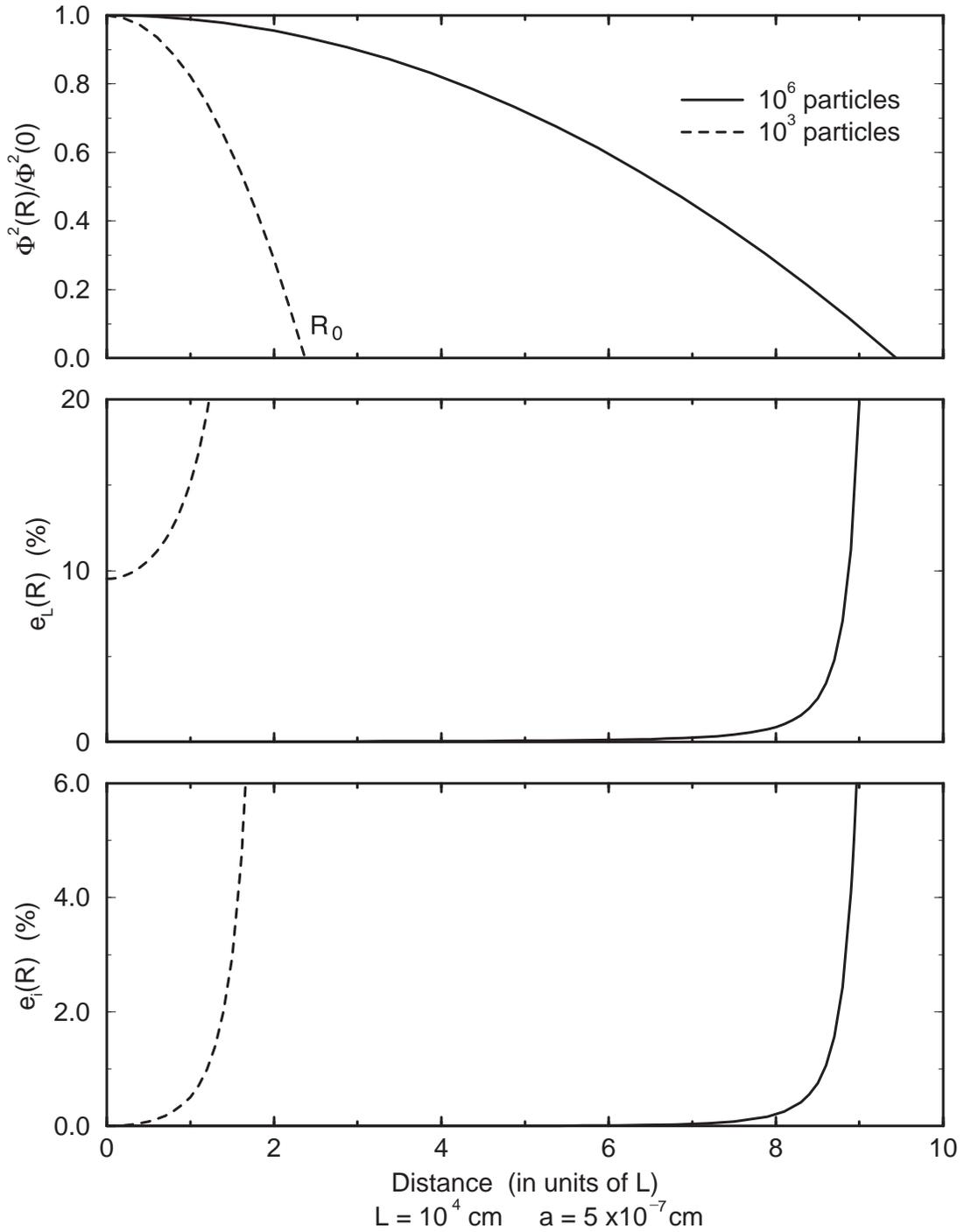 scaled 750}}
\caption{(a) Condensate wave functions for $N=10^3$ and $10^6$;
(b) Error incurred in neglecting kinetic term in NLSE;
(c) Error incurred in Thomas-Fermi approximation.
Length scale on horizontal axis is in units of $L$,
the extend of the groundstate wave function. Calculations are
done for $L=10^{-4}$ cm, scattering length=$5\times 10^{-7}$ cm.}
\end{figure}
\medskip

\section{Quasiparticle Excitation}

The local excitation energy should be measured from the
chemical potential:
\be
\epsilon_p(r)=\mu+\sqrt{\left[{p^2\over 2m}+\mu_{\rm eff}(r)\right]^2
-\mu_{\rm eff}^2(r)}
\ee
where 
\be
\mu_{\rm eff}(r)=\mu-V_{\rm ext}(r)
\ee
It describes a phonon with a position-dependent sound velocity. 
The excitation energy density of the system is given by
\be
g(\epsilon)=\sum_i\delta(\epsilon-\epsilon_i)
\ee
where $\epsilon_i$ is the energy of the $i$th excited state. In the
spirit of the Thomas-Fermi approximation, we take
\be
g(\epsilon)=\int{d^3r d^3p\over h^3}\delta(\epsilon-\epsilon_p(r))
\ee
The results for $N=10^3$ and $10^6$ are shown in Fig.6 and Fig.7,
with comparison to the ideal gas.

\medskip
\begin{figure}[p]
\centerline{\BoxedEPSF{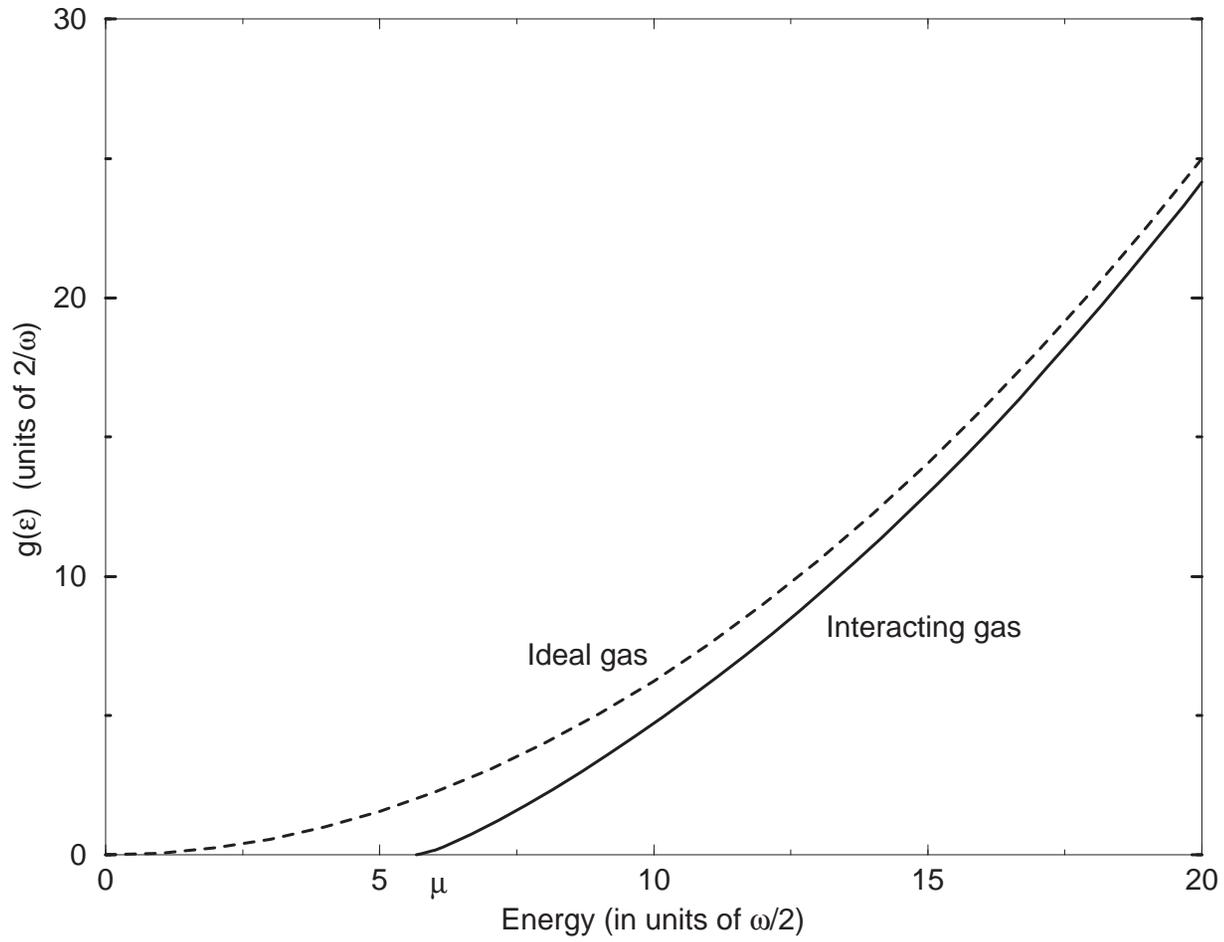 scaled 750}}
\caption{Density of states in harmonic trap, for $N=10^3$.}
\end{figure}
\medskip

\medskip
\begin{figure}[p]
\centerline{\BoxedEPSF{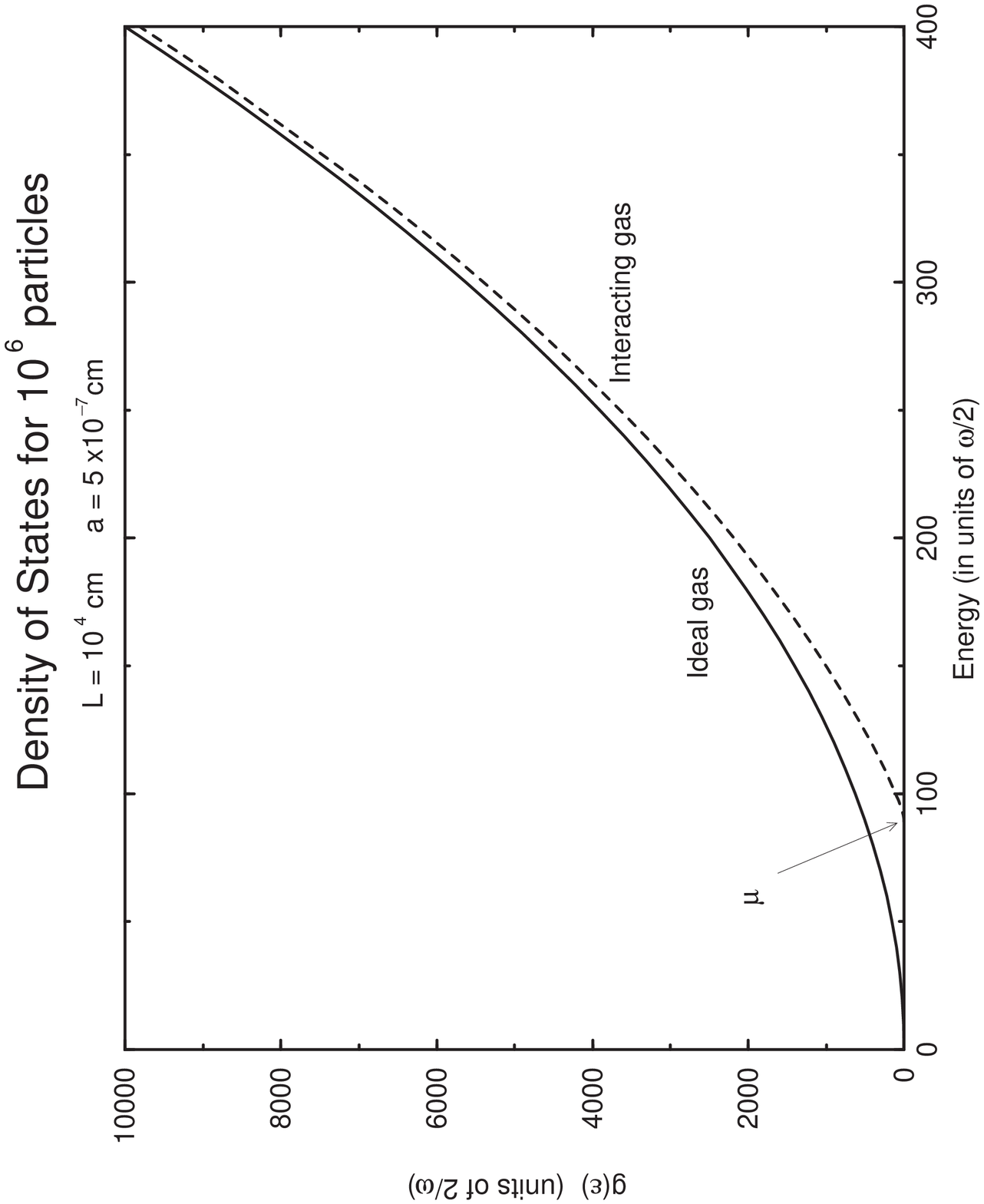 scaled 750}}
\caption{Same as Fig.6, but for $N=10^6$.}
\end{figure}
\medskip

Further details can be found in \cite{TTH}.

This work was supported in part by funds provided by 
the U.S. Department of Energy under cooperative agreement 
\# DE-FC02-94ER40818.

\end{document}